\documentclass[useAMS,usenatbib]{mn2e}
\usepackage{graphicx}

\newcommand{\e}{\mbox{e}^}
\newcommand{\uchII}{UC\,H\,{\small II}}
\newcommand{\ghz}{\,GHz}

\title[Fourier Transform Spectroscopy of HMC submm continuum emission]{Fourier Transform Spectroscopy of the submillimetre continuum emission from hot molecular cores}
\author[Friesen et al.]{R. K. Friesen$^{1,2}$\thanks{E-mail: rfriesen@uvastro.phys.uvic.ca}, D. Johnstone$^{2,1}$, D. A. Naylor$^{3}$ and G. R. Davis$^{4}$\\
$^{1}$Department of Physics and Astronomy, University of Victoria, Victoria BC  V8P 1A1, Canada\\
$^{2}$National Research Council of Canada, Herzberg Institute of Astrophysics, 5071 West Saanich Road, Victoria BC  V9E 2E7, Canada\\
$^{3}$Department of Physics, University of Lethbridge, Lethbridge AB T1K 3M4, Canada\\
$^{4}$Joint Astronomy Centre, Hilo, HI  96720, USA}

\begin{document}

\date{Accepted . Received }

\pagerange{\pageref{firstpage}--\pageref{lastpage}} \pubyear{2005}

\maketitle

\label{firstpage}

\begin{abstract}
We have used a Fourier Transform Spectrometer on the James Clerk Maxwell Telescope to study the submillimetre continuum emission from dust in three hot molecular cores (HMC). The spectral index $\beta$ of the dust emission for these sources has been determined solely within the 30\,GHz wide 350\,GHz (850\,\micron) passband to an accuracy comparable to those determined through multi-wavelength observations. We find an average $\beta\simeq1.6$, in agreement with spectral indices determined from previous submillimetre observations of these sources and with those determined for HMC in general. The largest single source of uncertainty in these results is the variability of the atmosphere at 350\,GHz, and with better sky subtraction techniques we show that the dust spectral index can clearly be determined within one passband to high accuracy with a submillimetre FTS. Using an imaging FTS on SCUBA-2, the next generation wide-field submillimetre camera currently under development to replace SCUBA at the JCMT in 2006, we calculate that at 350\,GHz it will be possible to determine $\beta$ to $\pm0.1$ for sources as faint as 400\,mJy/beam and to $\pm0.3$ for sources as faint as 140\,mJy/beam.
\end{abstract}


\section{Introduction}

Stars form in dense cores of dust and gas within molecular clouds, but remain cocooned in the material of their natal envelopes. Though young stars and protostars can be extremely luminous, the enveloping dust absorbs nearly all of the ultraviolet and optical starlight and reradiates it at longer wavelengths. In the submillimetre regime, the dust emission is generally optically thin. Submillimetre observations thus sample all the dust emission within the cloud along the line of sight, and can be used to determine the column density and mass of material present \citep{hild83}. Interpretation of the observations, however, depends on the assumed dust grain temperature, composition and size distribution. The effects of composition and grain size on the dust emission are usually combined into a dust opacity, $\kappa_\nu$, which can be reasonably fit empirically with a power law dependence on frequency in the submillimetre.

\begin{table*}
\centering
\begin{minipage}{106mm}
\caption{Positions and peak brightnesses for observed objects}
\label{positions}
\begin{tabular}{cccccc}
\hline
Source & $\alpha_{J2000}$ & $\delta_{J2000}$ & $I_{450}$ & $I_{850}$ & Distance\\
--- & \fh\,\fm\,\fs & \degr\,\,\,\arcmin\,\,\arcsec & Jy/8.5\arcsec~beam & Jy/15\arcsec~beam & kpc\\
\hline
G10.47 & 18 08 38.2 & $-$19 51 50 & 260 $\pm$ 65 & 39.0 $\pm$ 5.9 & 5.8\\
G31.41 & 18 47 34.5 & $-$01 12 43 & 130 $\pm$ 32 & 24.2 $\pm$ 3.6 & 7.9\\
G12.21 & 18 12 39.7 & $-$18 24 20 & 121 $\pm$ 30 & 13.0 $\pm$ 2.0 & 13.5\\
\hline
\end{tabular}

\medskip
The positions, peak brightnesses at 450\,\micron~and 850\,\micron, and distances to each of the hot molecular cores observed in this study, ordered by $I_{850}$. Positions and distances are taken from \citet{hatchell00}, while peak brightnesses were calculated from the archived SCUBA data.
\end{minipage}
\end{table*}

At submillimetre wavelengths, dust is generally assumed to radiate as a blackbody at some average temperature $T_d$ modified by the optical depth, $\tau_\nu \propto \kappa_\nu$. For optically thin emission, we then find $S_{\nu} \propto \kappa_\nu \, B_{\nu}(T_d)$. Assuming a power law dependence on frequency, the dust opacity $\kappa_\nu$ can be parameterized for a given opacity $\kappa_0$ at frequency $\nu_0$ as $\kappa_\nu=\kappa_0 (\nu / \nu_0)^\beta$, where $\beta$ is the spectral index of the dust emission. For optically thin emission, the dust continuum emission then follows the power law:
\begin{equation}
S_{\nu} \propto \kappa_{0}(\frac{\nu}{\nu_0})^{\beta}B_{\nu}(T_d)
\label{eqn:dust_law}
\end{equation}
Through mathematical modelling, $\kappa_\nu$ and $\beta$ have been calculated for various dust grain compositions and size distributions \citep{draine84,ossen94,poll94}, which can be compared with astronomical observations. Often, however, various combinations of the dust parameters can reproduce existing observational data with similar accuracy, while dust temperatures and densities can vary dramatically within star forming environments. Calculations often assume $\beta\simeq2$ at these wavelengths, based on work by \citet{hild83}. In practice, however, theoretical dust grain models find spectral indices anywhere between zero and three (see the summary by Goldsmith, Bergin \& Lis (1997))\nocite{gold97}, depending on the physical dust properties of the model, while observations have found a similar range of values in different states of the interstellar medium.

In high mass star forming regions and the cores containing warm protostars, \citet{will04} find the average frequency dependence of dust emission from more than $60$ high mass protostellar objects to be characterized by $\beta=0.9\pm0.4$. This is comparable to values observed in less evolved, colder objects \citep{gold97,visser98,hoger00,beuther04}, but is systematically lower than spectral indices determined through submillimetre observations of hot molecular cores (HMC) and ultra-compact H\,{\small II} (\uchII) regions, which are likely the next phases in the evolution of high mass protostars. In these more evolved cores, $\beta$ values have been determined to be near to \citep{osorio99} or slightly higher than \citep{hunter98} $\sim2$. 

The changing values of $\beta$ in these studies are indicative of the evolution of the dust itself in the changing physical environments associated with the process of star formation. The growth of dust grains, the formation of icy mantles, and changes in grain composition are likely all significant factors in the variation of the frequency dependence of the dust emission in these regions. In order to understand the properties and evolution of dust in star forming regions, observational methods are required that can discern between temperature and density effects and actual changes in the dust composition and opacity. 

To determine $\kappa_0$, observations must be calibrated against known dust column densities, which generally requires comparison with observations at shorter wavelengths. In contrast, the spectral index $\beta$ of the dust opacity can be calculated solely in the submillimetre by comparing observations at two widely separated submillimetre wavelengths. This method provides great leverage on $\beta$, but is complicated by calibration issues from observing in the different passbands, and has the disadvantage that the value determined for $\beta$ is dependent on the assumed dust temperature. The Planck function at 450\,\micron~is far from the Rayleigh-Jeans limit for dust temperatures less than $\sim40$\,K, and even at the higher temperatures expected in high mass star forming regions, the change in emission between passbands could be due to either small changes in temperature or large changes in the dust emissivity. The uncertainties in the determination of $\beta$ are thus large enough that the subsequently derived physical properties of dusty systems, such as the mass estimates of the material in dense cores in molecular clouds, may range over factors of a few. For this reason, it is beneficial to observe in one passband only.

One instrument that is well-suited to this problem is a Fourier Transform Spectrometer (FTS). FTS have several advantages and disadvantages compared with other spectrometer designs. The highest spectral resolution of an FTS is determined by the maximum optical path difference between the two beams of the interferometer, which is set by the length of the translation stage. In practice, space limitations require that most submillimetre FTS provide low to intermediate spectral resolution (R\,$\sim200$ - R\,$\sim2000-3000$ at 345-375\ghz), excellent for observations of sources with wide, bright spectral lines and continuum emission. For these studies FTS have a distinct advantage over high spectral resolution but narrow bandwidth heterodyne receivers. The instantaneous bandwidth of an FTS is inherently broad, and in the submillimetre is limited only by the atmospheric transmission windows, enabling a wider spectral coverage within these windows than is possible with the heterodyne receivers currently in use. In the past decade, two groups have successfully used FTS to study the submillimetre emission from several astronomical sources, such as the Orion molecular cloud \citep{serabyn}, the Sun \citep{naylor00}, and the atmospheres of planets \citep{naylor94,davis97}.

A submillimetre FTS operated at low resolution can sample the continuum dust emission from a source with a large enough bandwidth that the change in dust emissivity with frequency, and thus $\beta$, can be determined solely within one passband. At very low temperatures, it is still difficult to separate the temperature component of the emission from the dust power-law emissivity function, but single bandpass observations are able to place limits on the combined dust temperature and $\beta$. 

In this study, we present low resolution 350\,GHz continuum observations of four hot molecular cores performed with a submillimetre FTS of the Mach-Zehnder design (MZ FTS) developed for use at the JCMT by the FTS group at the University of Lethbridge, AB, Canada (described in detail by \citet{naylor03}). 

We describe in the following the observational procedure and analysis of the data. We compare the dust spectral index values determined for our sources with those calculated from previous submillimetre measurements, and discuss the detrimental effects of the variable submillimetre atmosphere on our sensitivity. Finally, we comment on the continuum measurement capabilities of an imaging FTS currently in preparation for use at the JCMT with SCUBA-2.

\section{Observations}

\subsection{Source selection}

The sources for this pilot study were chosen from a sample of HMC from work by \citet{hatchell00} for their large submillimetre fluxes and lack of extended structure as shown by \citet{hatchell00} and \citet{walsh03}. The sources, G10.47, G12.21 and G31.41, have been previously studied in the submillimetre regime with the Submillimetre Common User Bolometer Array (SCUBA) at the JCMT \citep{hatchell00, walsh03} and in the radio continuum (\citet{wc89}, among others). All three have also been detected in numerous molecular lines and maser emission (see the detailed source descriptions in \citet{hatchell00}). Table \ref{positions} gives source positions and distances from \citet{hatchell00}, and 450\,\micron~(660\,GHz) and 850\,\micron~(350\,GHz) peak brightnesses calculated using the archived SCUBA data\footnote{Guest User, Canadian Astronomy Data Centre, which is operated by the Dominion Astrophysical Observatory for the National Research Council of Canada's Herzberg Institute of Astrophysics}. Each source is associated with one or more \uchII~regions near the peak of the submillimetre emission. The hot core in G10.47 contains three \uchII~regions, G10.47-0.03A, B and C, while the other sources are coincident with one \uchII~region each, G12.21-0.10 and G31.41+0.31. These HMC are very peaked in the 850\,\micron~emission, although G31.41 does contain some more extended emission, and the hot cores themselves of these sources are much smaller than the JCMT beam with diameters of only 1-2\arcsec. Core mass estimates are on the order of a few thousand M$_\odot$ \citep{hatchell00}. A fourth HMC, G43.89, was observed by us to only S/N\,$\sim1.5$, too low to analyse with any degree of certainty, and is therefore omitted from further discussion. Observations of Mars were also performed for flux calibration.

\subsection{Observational method}
\begin{figure}
\includegraphics[width=84mm]{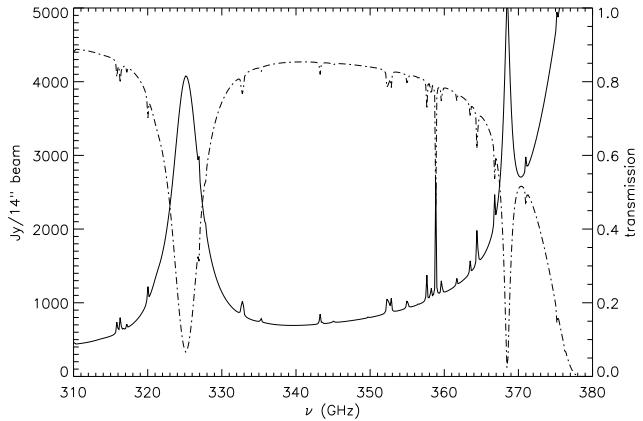}
\caption{The atmospheric radiance (solid line) in units of Jy/15\arcsec~beam and transmission (dash-dotted line) calculated for precipitable water vapour levels of 1.0mm.}
\label{fig:ULTRAM}
\end{figure}

The observations were performed during 2003 April 21 - 28, using the MZ FTS mounted on the right Nasmyth platform of the JCMT on Mauna Kea, Hawaii. The April 2003 weather was good to excellent with the zenith opacity measured at 225\ghz~(1.3\,mm), $\tau_{225}$, $\sim0.07$ for most of the observations, giving precipitable water vapour (pwv) levels averaging $\sim1.1$\,mm. At 350\,GHz, the FTS bandwidth is $\sim30$\ghz. The FTS was operated at low spectral resolution (1.5\ghz), resulting in $\sim\,20$ resolution elements across the band. The full width half maximum (FWHM) of the 350\,GHz~beam at the JCMT is $\sim15\arcsec$, which was matched by a cold (4\,K) aperture located at the field stop of the detector system \citep{naylor99}. The atmospheric emission and transmission window at  350\,GHz is shown in Figure \ref{fig:ULTRAM}. 

Spectra were obtained in groups, with five spectra taken of a molecular core immediately followed by five spectra of the background sky, +1425\arcsec~in right ascension from the source. The offset in right ascension was calculated to compensate for the change in airmass of the source due to the Earth's rotation during the time taken to obtain five spectra, allowing the sky measurements to be taken at the same airmass as the original source observations. This is crucial for accurate background sky subtraction, as the brightness and variability of the atmosphere at 350\,GHz is easily the largest source of uncertainty in our final results. Five scans were obtained approximately every 90 seconds. Each target source was observed for between three and four hours, with approximately half of that time spent on source and half spent on background observations. Due to its brightness, Mars was observed for a half hour. A pointing check performed approximately every 45 minutes resulted in small ($\sim1-2\arcsec$) corrections to the pointing. 

\subsection{Data reduction}

\begin{figure*}
\begin{minipage}{170mm}
\begin{center}
\includegraphics[width=84mm]{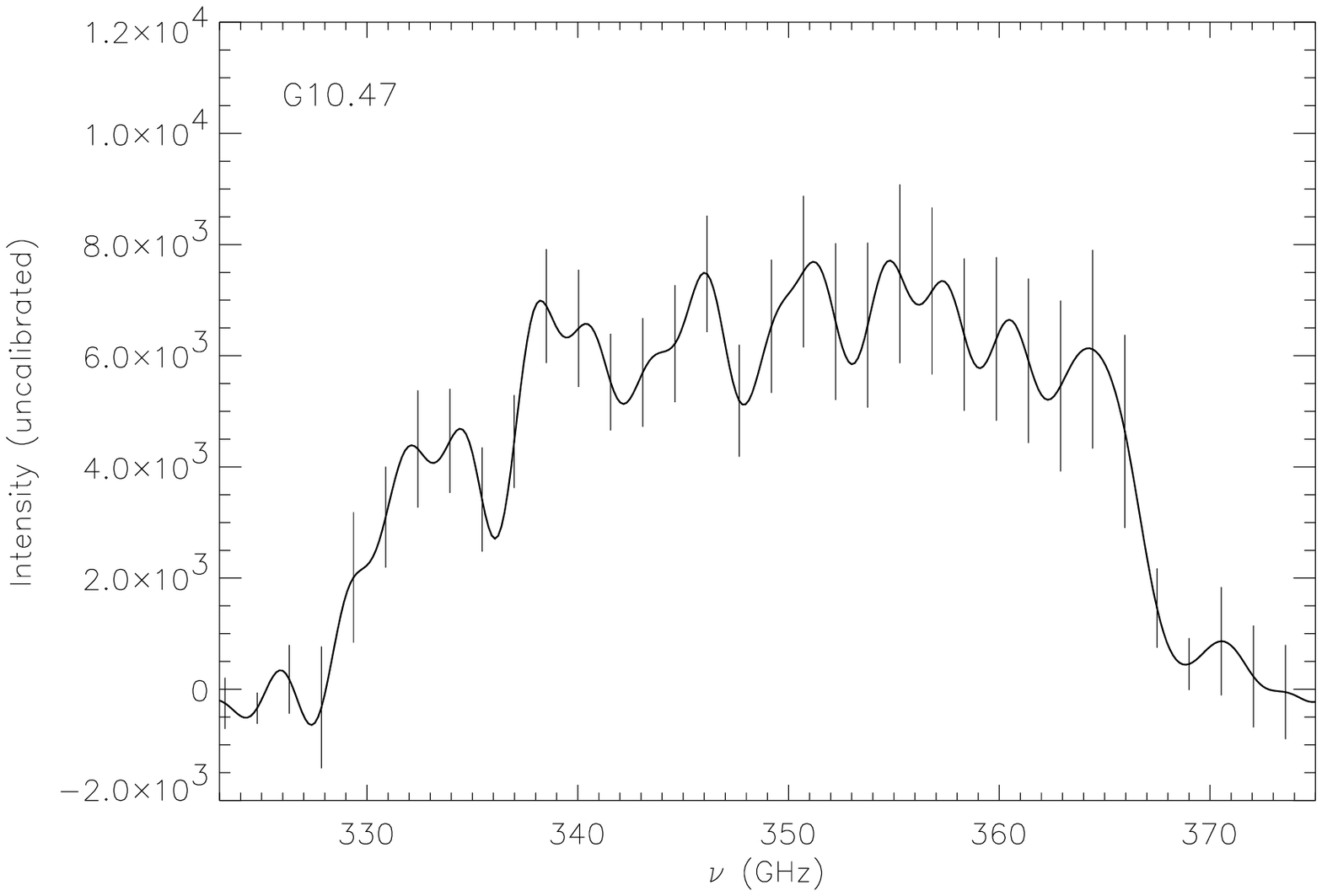}
\hfill
\includegraphics[width=84mm]{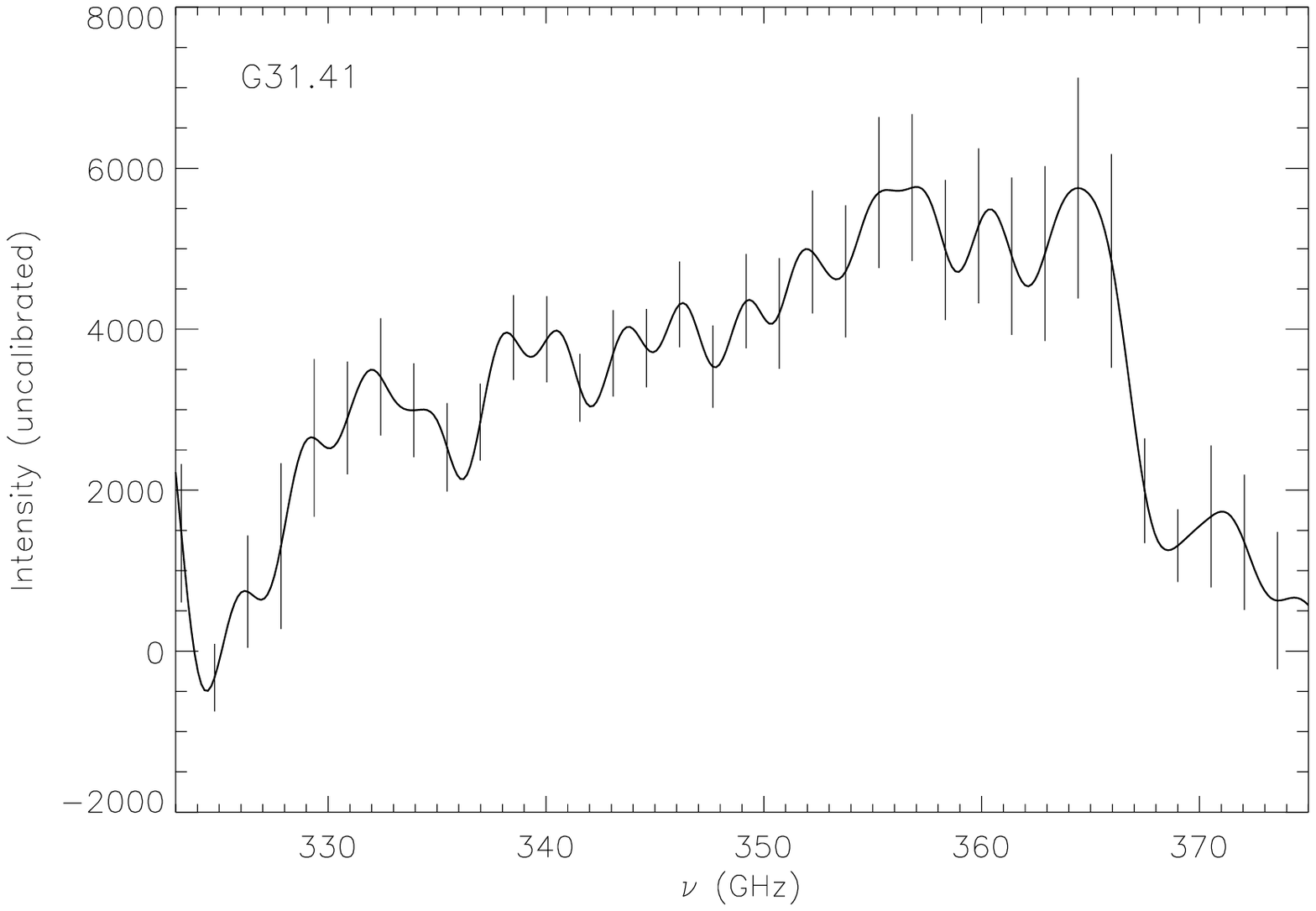}\\
\includegraphics[width=84mm]{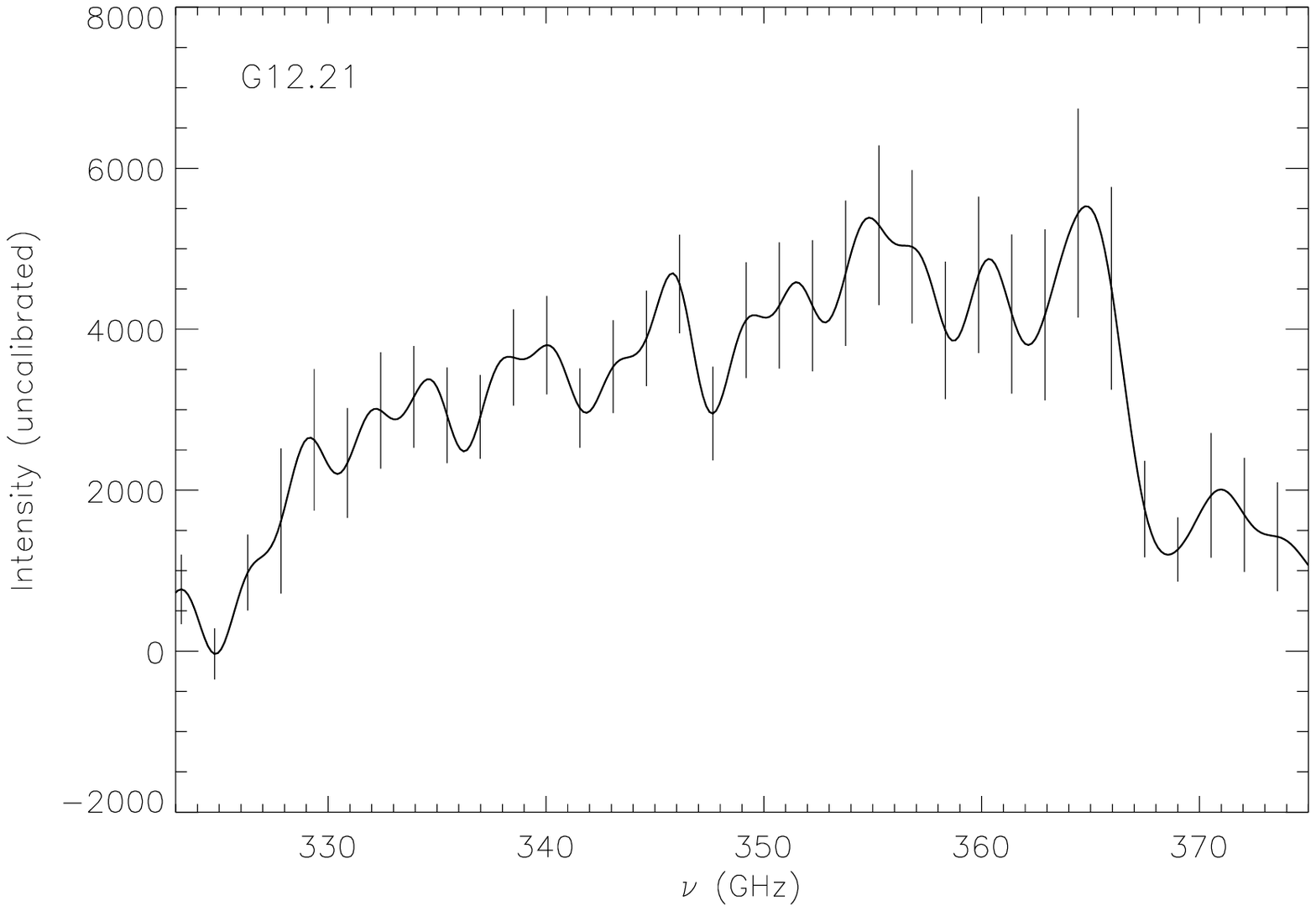}
\hfill
\includegraphics[width=84mm]{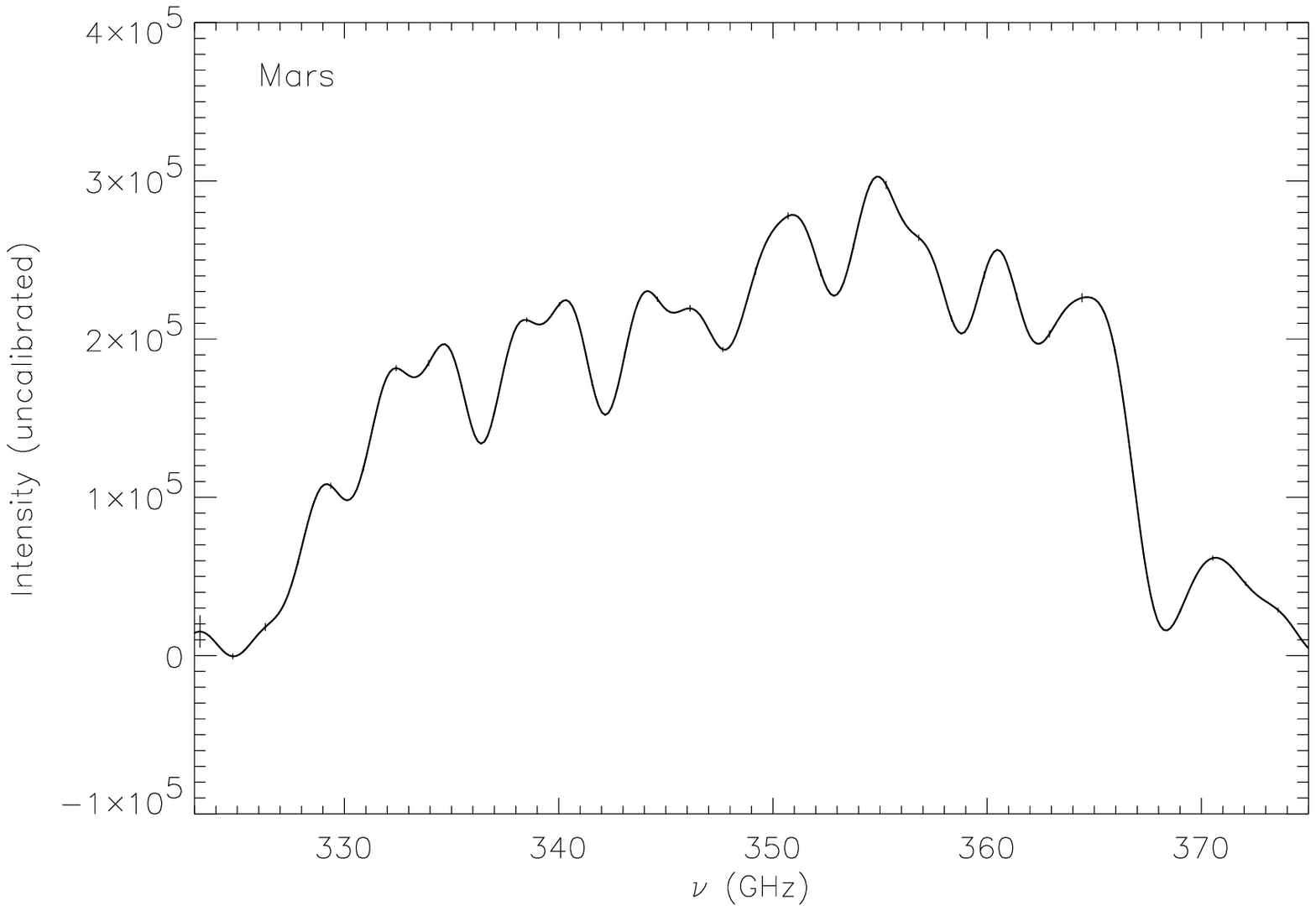}
\caption{The raw, coadded, sky-subtracted spectra of G10.47, G12.21, G31.41 and Mars before calibration or atmospheric transmission correction. Error bars are at the 1\,$\sigma$ level. Note that the error bars for Mars lie within the line thickness.}
\label{fig:mean_spc}
\end{center}
\end{minipage}
\end{figure*}

Preliminary data reduction was performed using processing pipeline software written for the FTS \citep{naylor03}. The data were first manually screened for transients, such as cosmic rays, which occur approximately once in every ten interferograms and produce large intensity spikes at one or two data points in the interferograms. These spikes are easily detected and can be removed within the pipeline. Some spikes in the interferograms were spread over multiple data points; these interferograms were omitted from the final dataset. Only a handful of interferograms were rejected for this reason. Since the FTS optics and detector have negligible dispersion over the range of interest, a linear phase correction was applied to each interferogram. The individual interferograms were then Fourier transformed. The spectra were then coadded into the on- and off-source groups of five, and the coadded off-source background scans were subtracted from the coadded on-source scans. The coadded, sky-subtracted spectra of G10.47, G12.21, G31.41 and Mars are shown in Figure \ref{fig:mean_spc}. 

\begin{figure}
\includegraphics[width=84mm]{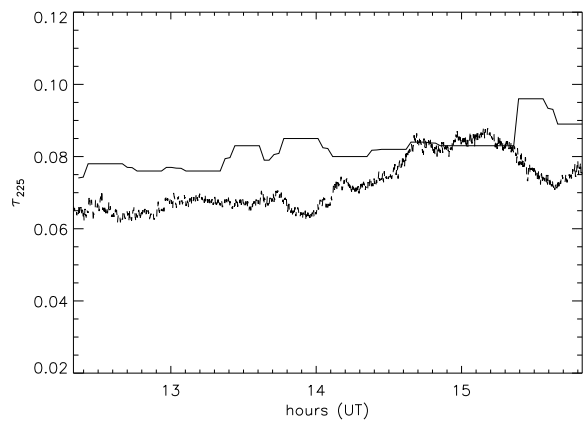}
\caption{The opacity at 225\,GHz measured from the CSO and determined through measurements of the precipitable water vapour by the JCMT's water vapour monitor (WVM) during observations of G10.47. The solid line is the CSO-measured opacity, corrected to match the JCMT observational airmass, which operates at fixed azimuth and updates approximately every 10-15 minutes. The WVM is pointed slightly off the main JCMT beam, and updates every 6 seconds.}
\label{fig:pwv_vs_cso}
\end{figure}

Each background-subtracted spectrum was then corrected for atmospheric transmittance within the band using the technique described by \citet{davis97}. Updated atmospheric opacities were calculated using the University of Lethbridge Transmission and Radiance Atmospheric Model (ULTRAM; \citet{chapman}). ULTRAM is a radiative transfer atmospheric model which calculates the radiance and transmission of the atmosphere through a line-by-line, layer-by-layer analysis, and was written with the specific goal of accurately modelling the atmosphere above Mauna Kea. The model parameterizes the atmospheric transmittance in terms of the pwv and airmass, and includes H$_2$O, O$_3$ and O$_2$ as atmospheric absorbers. The airmass of each scan was recorded with each interferogram. The opacity at 225\ghz, $\tau_{225}$, was recorded by a dipping radiometer operated by the Caltech Submillimeter Observatory (CSO). The use of this opacity value in the data analysis introduces some uncertainty into the final results, as the CSO radiometer operates at a fixed azimuth which rarely corresponds with the azimuth of the JCMT observations, and only updates approximately every 10-15 minutes, while the pwv levels in the atmosphere may vary significantly on shorter timescales. This can be seen in Figure \ref{fig:pwv_vs_cso}, where $\tau_{225}$ from the CSO is plotted against $\tau_{225}$ determined from measurements of the pwv recorded by the JCMT's water vapour monitor (which is pointed slightly off the main JCMT beam) during the observations of G10.47. The CSO $\tau_{225}$ value is often offset from that recorded by the water vapour monitor, and does not reflect the short timescale of the opacity variability. The water vapour monitor, while operating during our observations, was new to the telescope and thus was not incorporated into the data reduction pipeline. In future, the use of the water vapour monitor will provide more accurate atmospheric pwv measurements.

Mars observations were then used to calibrate the final, atmospheric transmission corrected, coadded spectra.

\section{Analysis}

\subsection{Source signal analysis}

We first discuss sky subtraction and instrumental contributions to the FTS signal. We then detail calibration issues, which include atmospheric variation and spectral line contamination of the continuum emission. 

The measured signal $S$ as a function of frequency $\nu$ from an astronomical source through the FTS can be expressed as
\begin{eqnarray}
S(\nu) &=& R(\nu)\,G\,[\eta_a\,J_s(\nu)\,\e{-\tau(\nu)} + \eta_a\,J_{sky}(\nu) (1-\e{-\tau(\nu)}) \nonumber \\
&+& (1-\eta_a)\,J_{amb}(\nu)-J^\prime(\nu)] 
\end{eqnarray}
where the contribution from any additional background has been neglected. Here, $\tau(\nu)$ is the atmospheric optical depth as a function of frequency $\nu$, the telescope aperture efficiency is given by $\eta_a$,  $R(\nu)$ is the responsivity or response function of the FTS, and $G$ is the detector gain when observing an astronomical source. For all observations, the gain was set to $G=10^4$. $J_s(\nu)$ is the true spectrum of the source, $J_{sky}(\nu)$ is the true spectrum of the atmosphere, $J_{amb}(\nu)$ is the power received from ambient temperature surfaces of the telescope, and $J^\prime(\nu)$ is the power received through the $2^{nd}$ port of the FTS, which is differenced in the interferometric measurement \citep{naylor00}. The first term therefore represents the power received from the astronomical source itself through the atmosphere and instrument, the second term represents the power received from the atmosphere and the third term represents the power received from the surfaces of the telescope, which are at ambient temperature.

Observations at the background position produce solely the intensity received from the sky, $S_{sky}(\nu)$, and ambient surfaces:
\begin{eqnarray}
S_{sky}(\nu) &=& G\,R(\nu)\,[\eta_a\,J_{sky}(\nu)(1-\e{-\tau(\nu)}) \nonumber \\
&+& (1-\eta_a)J_{amb}(\nu)-J^\prime(\nu)]
\end{eqnarray}
If the atmospheric temperature and opacity are assumed to remain unchanged during each pair of source and background scans, the difference of the on- and off-source scans gives the observed power of the source itself, $S_s(\nu)$:
\begin{eqnarray}
S_s(\nu)&=& S(\nu) - S_{sky}(\nu) = R(\nu)\,G\,\eta_a\,J_s(\nu)\,\e{-\tau_s(\nu)}
\end{eqnarray}
Sky brightnesses at 345\ghz~for the typical pwv levels during our observations of $\sim1.1$\,mm are $\sim800$\,Jy/beam, an order of magnitude larger than our brightest source, making accurate background subtraction essential. During the observations, measurements of the precipitable water vapour, from the CSO radiometer, and observational airmass were recorded. With this information, the atmospheric model described in Section 2 was used to determine the transmission of the atmosphere in the direction of the source, and thus the contribution of the term $\e{-\tau(\nu)}$ to the observed source intensity $S_s(\nu)$. Typical values of $\tau_{225}$ during the observations were $\sim0.07$, leading to atmospheric transmission levels at 345\ghz~of $\sim75$\%.

Observations of Mars and a brightness temperature model were used for calibration of the spectra obtained by the instrument. At the time of our observations, Mars had an angular diameter of 8.97\arcsec. The Mars observations were done in the same on- and off-source pattern as the HMC observations, and the intensity $S_m(\nu)$ received from Mars after sky subtraction is thus
\begin{eqnarray}
S_m(\nu)&=& S(\nu) - S_{sky}(\nu) = R(\nu)\,G\,\eta_a\,J_m(\nu)\,\e{-\tau_m(\nu)}
\end{eqnarray}

The brightness temperature of Mars, $T_{B,m}$, on the date of our observations was calculated using the STARLINK FLUXES program \citep{starlink}. The Mars flux, $J_m(\nu)$, at the telescope can then be calculated:
\begin{equation}
J_m(\nu) = \frac{2\nu^2k}{c^2T_{B,m}}
\label{eqn:flux}
\end{equation}

The true spectrum of the hot core sources, $J_s(\nu)$, can then be determined: 
\begin{equation}
J_s(\nu)=\frac{S_s(\nu)\,\e{-\tau_m(\nu)}}{S_m(\nu)\,\e{-\tau_s(\nu)}} J_m(\nu)
\end{equation}

\begin{figure}
\includegraphics[width=84mm]{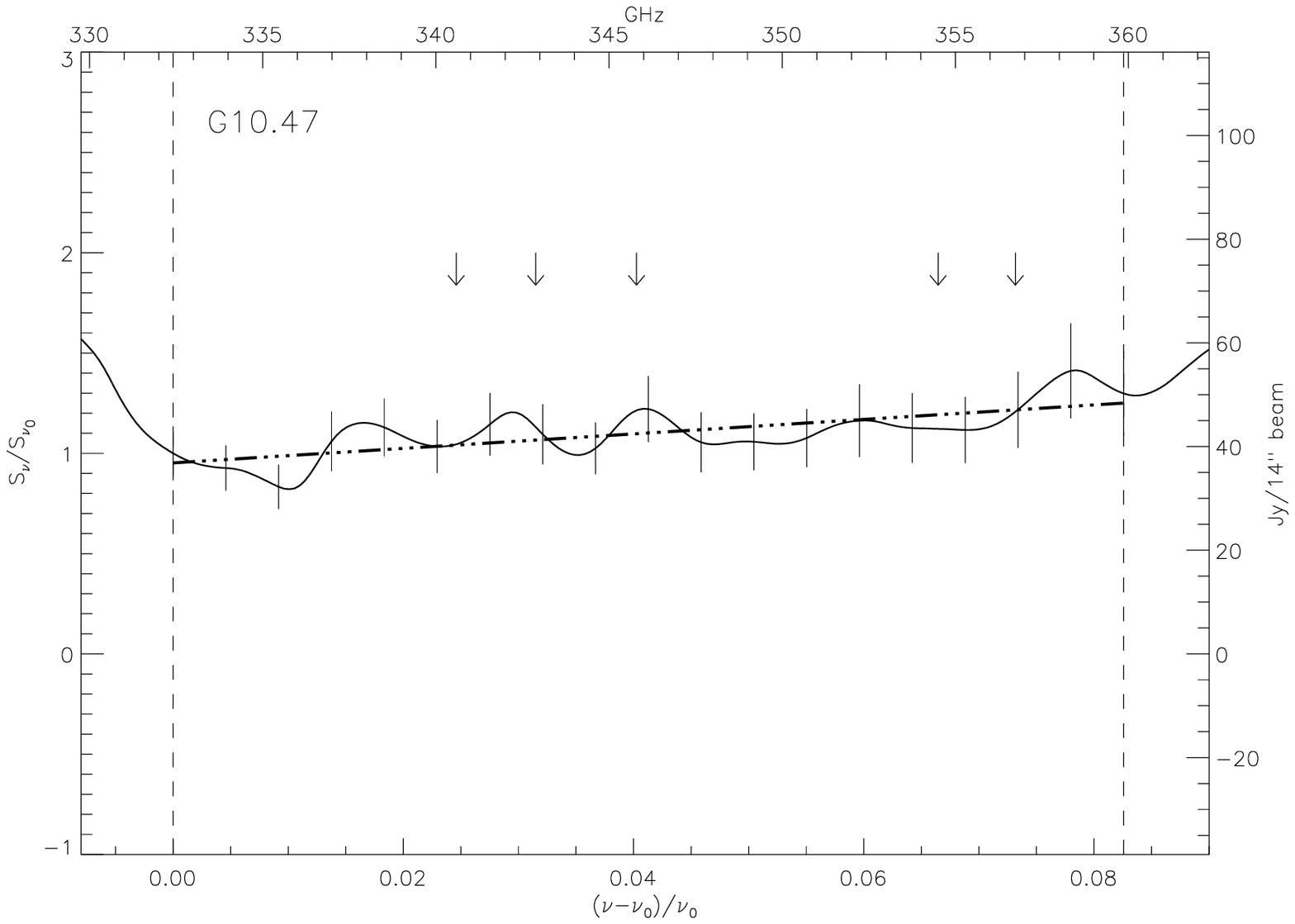}\\
\includegraphics[width=84mm]{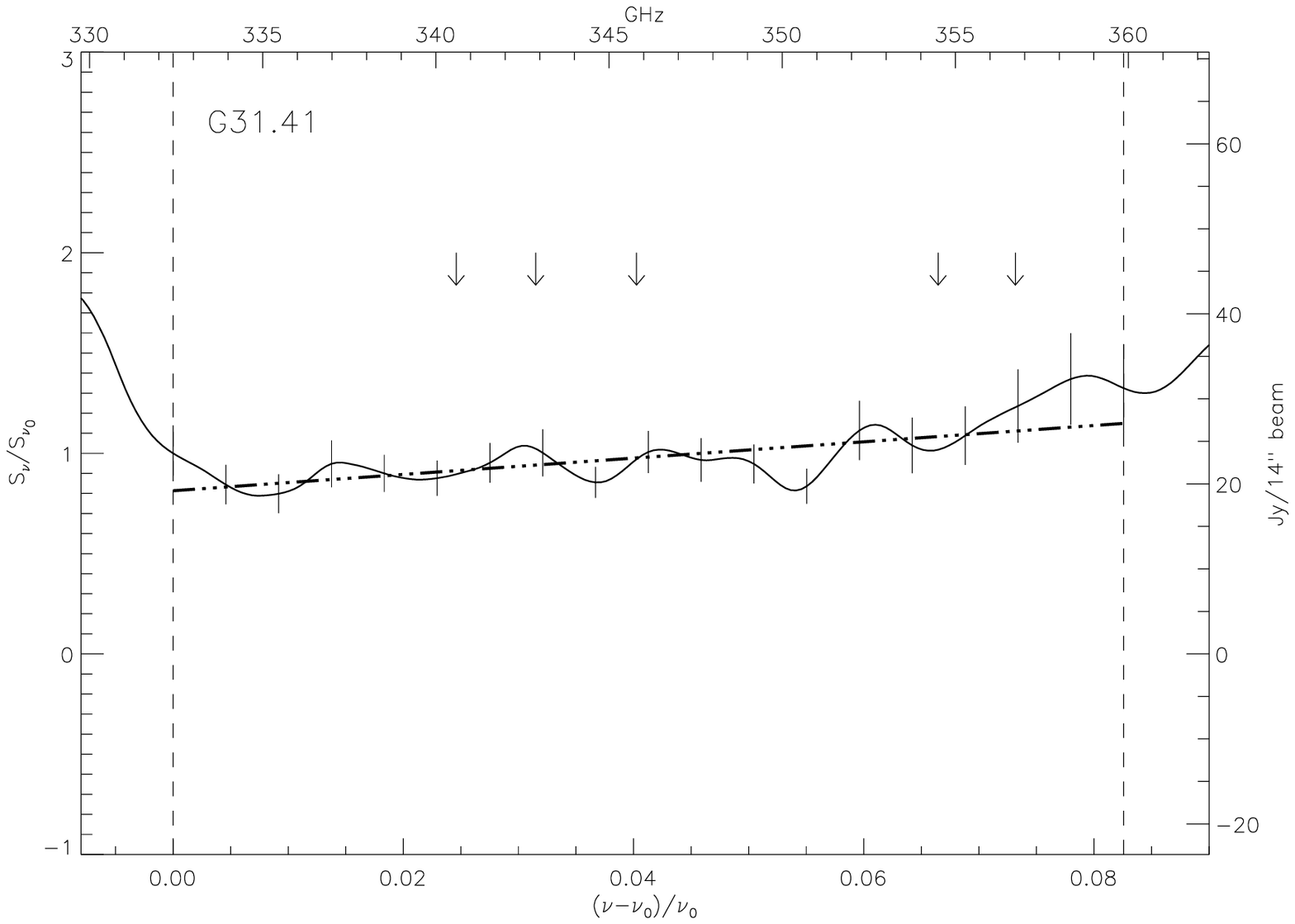}\\
\includegraphics[width=84mm]{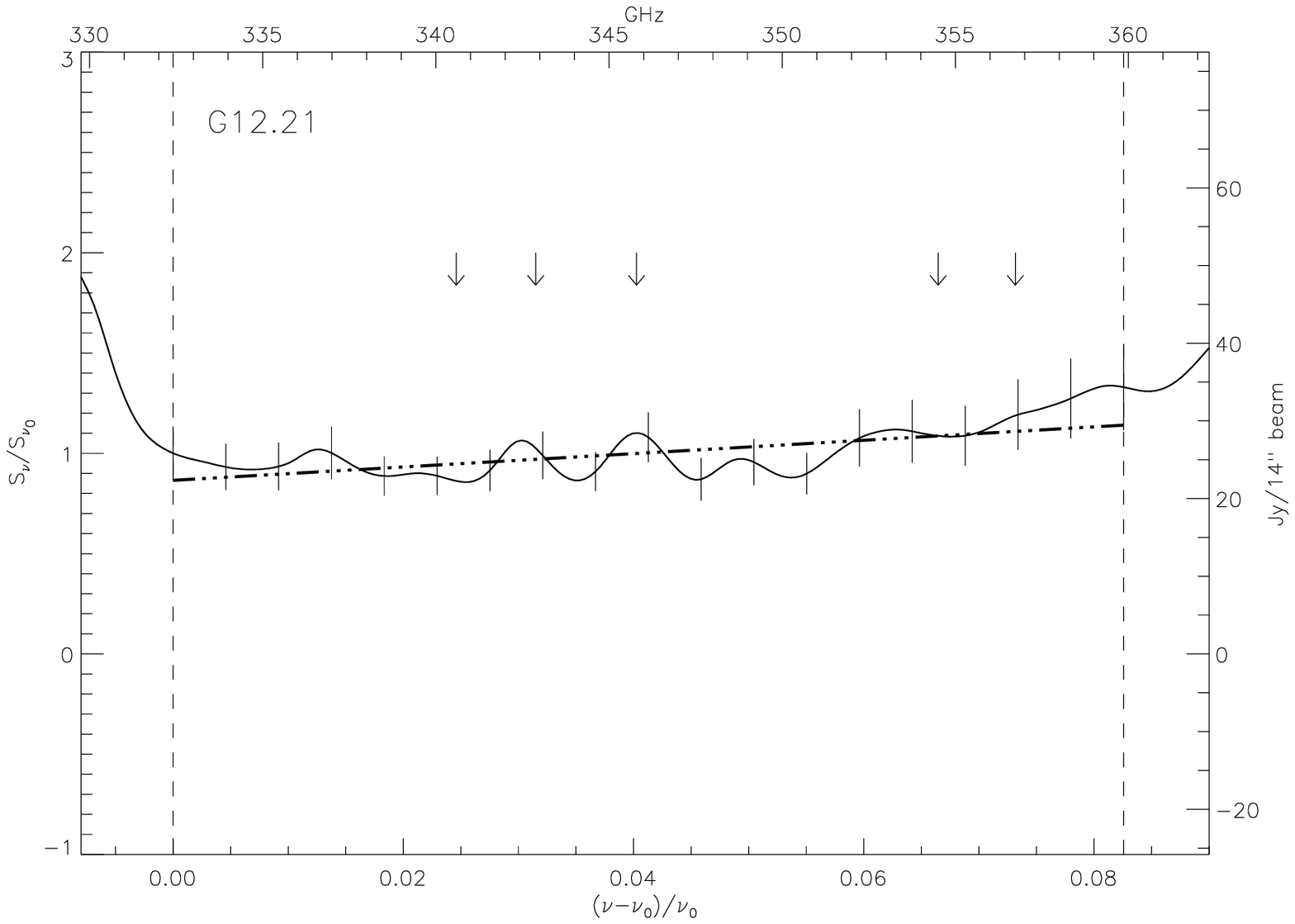}
\caption{Final calibrated spectra of the hot core sources (top to bottom: G10.47, G31.41, G12.21) with 1$\sigma$ error bars at 1.5\ghz~resolution. The dash-dotted line is the best fit linear slope. The dashed lines indicate the fitting window. Arrows indicate the positions of five molecular emission lines, listed in the text, expected to be strong in HMC. With the exception of the CO(3-2) line at 345.6\,GHz, no spectral features are coincident with the expected bright line positions.} \ \ \ \ \
\label{fig:slopes}
\end{figure}

The final 350\,GHz spectrum of each source can be seen in Figure \ref{fig:slopes} in units of Jy\,/beam. The derived intensity near the edges of the band has larger error due to the strong water vapour transitions seen in Figure \ref{fig:ULTRAM}, which define the 350\,GHz spectral window. Variations in the column abundance of water vapour between source and background measurements result in greater noise at the steepest part of the water vapour line profile.

\subsection{Signal-to-noise and atmospheric effects}

The instrumental signal-to-noise (S/N) ratio was calculated from the MZ FTS characteristics in \citet{naylor03} for the spectral resolution and specific observing times for each source, and is solely due to the detector noise of the FTS. The S/N ratios of the final spectra are compared with the instrumental S/N ratios in Table \ref{SN}, with the average S/N achieved across the band a factor of order $\sim10$ lower than the instrumental value for all sources. With all detector effects accounted for in the instrumental S/N ratio, the sensitivity degradation is solely due to the atmospheric variance during our observations. 

\begin{table}
\caption{Instrumental and observed S/N values}
\label{SN}
\begin{tabular}{cccc}
\hline
Object & Time on source & S/N$_{Ins}$ & S/N$_{Obs}$ \\
--- & (hours) & --- & --- \\
\hline
G10.47 & $3.5$ & $147$ & $8.5$\\
G31.41 & $4.0$ & $99$ & $8.0$ \\
G12.21 & $4.0$ & $73$ & $7.4$ \\
\hline
\end{tabular}

\medskip
The instrumental and observed signal-to-noise ratio for each source.
\end{table}

At submillimetre wavelengths, the brightness and variability of the sky is an overwhelming obstacle to accurate flux calibration, and observations generally require some form of chopping between source and background measurements in order to remove the sky emission. Ideally, the time between source and background measurements is as short as possible to limit the level of sky variance between observations \citep{arch02}. While our observational method was designed to minimize the effects of the ever-changing atmosphere on the data by switching between source and background observations every 90\,s, the atmosphere still contributed substantially to our uncertainties for several reasons. First, the opacity varies on timescales shorter than 90\,s, as shown in Figure \ref{fig:pwv_vs_cso}. Second, delays in slewing the telescope to the background position resulted in slight differences between the airmass of the background and source scans, introducing a small systematic effect by effectively offsetting the opacity in the background scan relative to the on-source scan. 

As a quick test, we can think of a small increase in opacity as a small increase in the precipitable water vapour (with the caveat that there are other minor contributions to the opacity). Using atmospheric emissivities calculated for various pwv levels with ULTRAM, we have calculated that a modest jump in pwv of 0.05\,mm, very reasonable when combining small deviations between source and background airmass with actual changes in the atmospheric pwv levels, increases the atmospheric emission by $\sim35$\,Jy/beam at 350\,GHz. The pwv values calculated from the CSO $\tau_{225}$ for the observations contain jumps of this size and a few even larger. For this reason, stable weather is more crucial to this study than particularly good weather. A change in atmospheric emission of 35\,Jy is comparable to the flux of our brightest source, G10.47, and brighter than the others (see Table \ref{positions}). As a consequence, some of the subtractions of the source and background spectra produced unphysical, negative flux measurements, while others resulted in impossibly high source brightnesses. When the spectra are coadded, however, on the whole these overly positive and negative scans will average out unless there is a large systematic offset between source and background observations or the pwv levels in the atmosphere are systematically increasing or decreasing. 

While there was a small systematic offset in the airmass between the on- and off-source scans, it was not large enough to produce significant changes in the sky emission except in the case of G31.41, which was observed while setting. For this source, the offset may have decreased the observed flux by at most 1-2\,Jy/beam. During observations of G10.47, the recorded atmospheric opacity steadily increased, although it remained low. After atmospheric transmission corrections, the recorded intensity of this source near the start of the observations was greater by $\sim10$\% on average than that recorded later. Again, this may have decreased the observed flux by a few Jy/beam at most, and is the largest contributor to the uncertainty per data point in the spectrum. There do not appear to be any systematic trends in the atmospheric opacity for the G12.21 observations.

\begin{figure}
\includegraphics[width=84mm]{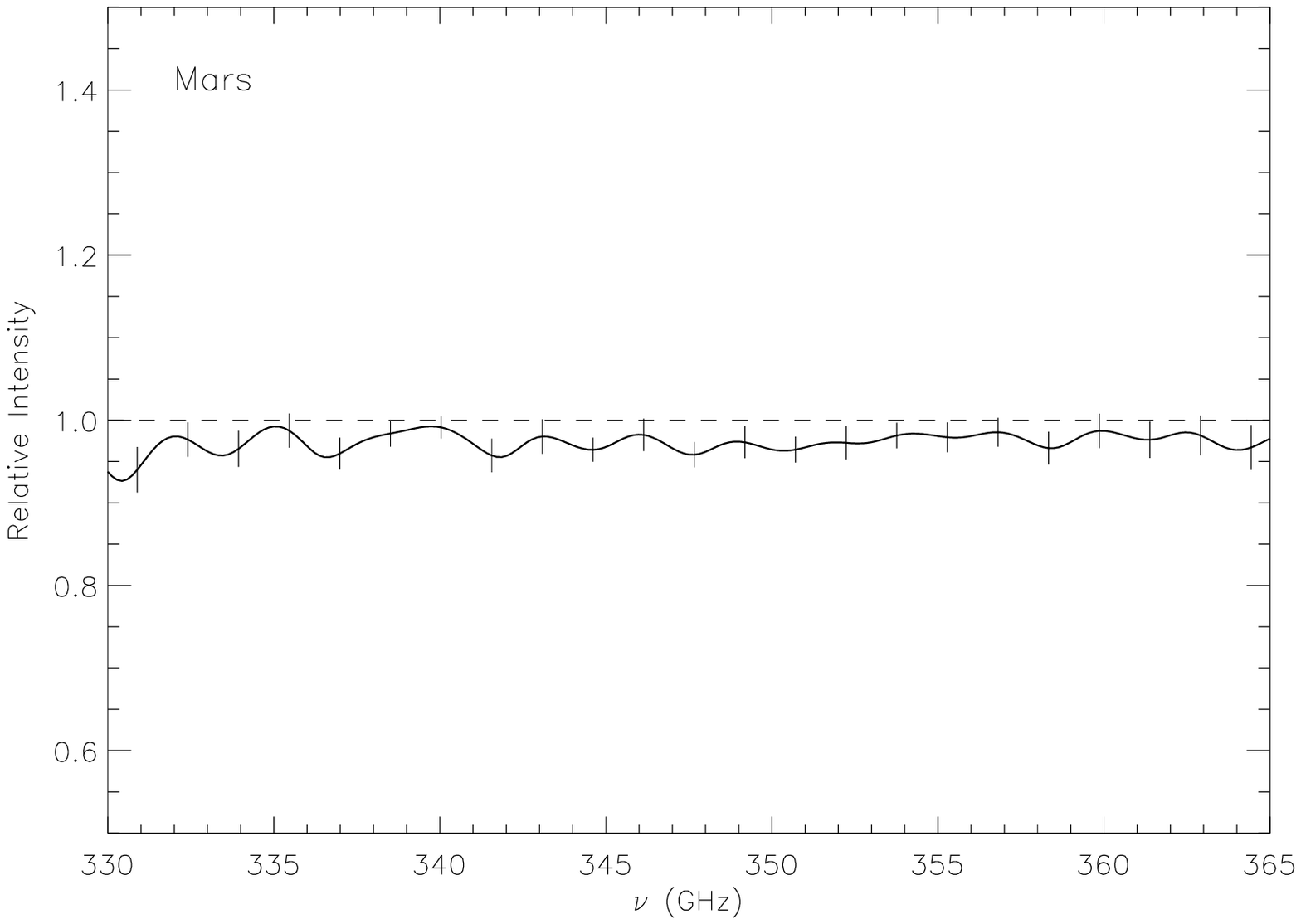}\\
\includegraphics[width=84mm]{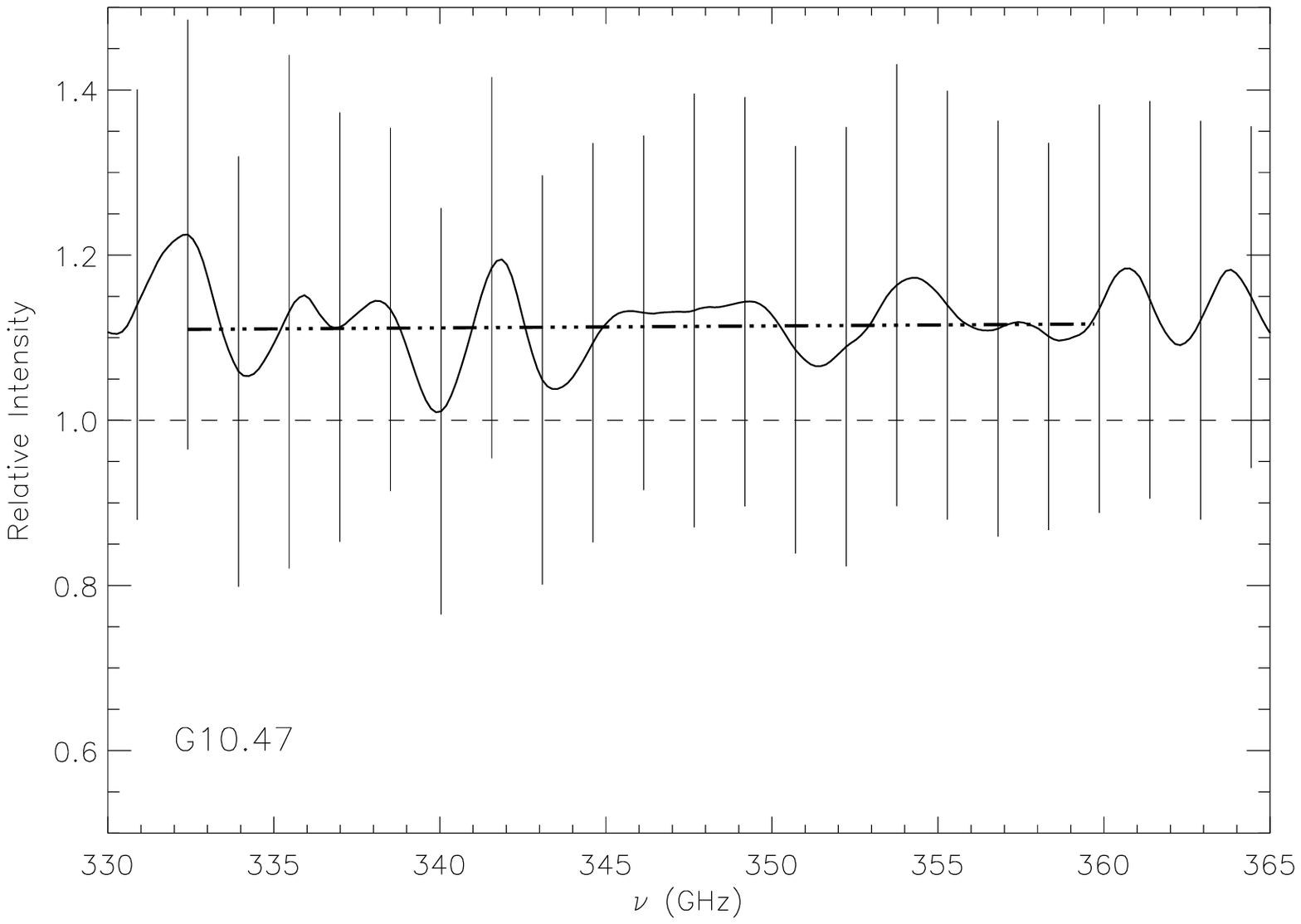}
\caption{Sky-subtracted spectra, divided into two groups corresponding to observations made in the first and second halves of each night, then coadded and ratioed for Mars and G10.47. The dash-dotted line indicates the best-fit slope of the ratioed data for G10.47, showing that while the variable atmosphere may introduce offsets into our final observed intensity of the sources, it has not significantly affected the slope of the continuum emission.}
\label{fig:source_vs_source}
\end{figure}

To estimate the effects of the varying atmosphere on our spectra, and to determine the level of any systematic effects which may have been introduced into the data, we divided the sky-subtracted spectra into two groups for each source corresponding to observations made in the first and second halves of each night. We then coadded the spectra in these groups, and ratioed the coadded spectrum of the first half with that of the second. Figure \ref{fig:source_vs_source} shows this ratio for Mars, our calibrator, and G10.47. It is clear that while there is a small shift in overall flux between the early and later observations, within the band there is little overall change in the continuum slope. The fainter sources showed similar results, and we thus expect that while the variable atmosphere may introduce offsets into our final observed intensity of the sources, it has not significantly affected the slope of the continuum emission. 

\subsection{Signal strength}

The final brightness values recorded by the FTS for each source are listed in Table \ref{fluxes} with calculated uncertainties. The uncertainty in the final flux value comes from the uncertainty in the observed flux of the source, $S_s(\nu)$, the observed flux of Mars, $S_m(\nu)$, the uncertainty in the Mars brightness temperature model, and the uncertainty in the atmospheric transmission, $\e{-\tau(\nu)}$. From the S/N of the observations, the uncertainty in $S_s(\nu)$ is $\sim11-12$\% for each source. We take an uncertainty in the atmospheric transmission of 10\%, while the estimated uncertainty for the Mars model is of order 5\%. The uncertainty in $S_m(\nu)$ is negligible compared to the other factors due to Mars' brightness. Added in quadrature, this gives an uncertainty in the calibrated source flux of 16\%. Additionally, the error lobe of the 850\,\micron~(350\,GHz) JCMT beam may contain up to $\sim5$\% of the source flux, making the final uncertainties of order 21\%. 

\begin{table}
\caption{Comparison of source brightnesses from SCUBA and the FTS}
\label{fluxes}
\begin{tabular}{ccc}
\hline
Object & $I_\nu$ (SCUBA) & $I_\nu$ (FTS) \\
--- & Jy/15\arcsec~beam & Jy/15\arcsec~beam \\
\hline
G10.47 & $39.0\pm5.9$ & $44.0\pm8.8$ \\
G31.41 & $24.2\pm3.6$ & $24.6\pm4.9$ \\
G12.21 & $13.0\pm2.0$ & $27.2\pm5.4$ \\
\hline
\end{tabular}

\medskip
A comparison of the 850\,\micron~source brightnesses determined using the FTS with the peak brightnesses recorded by the Submillimetre Common User Bolometer Array. The uncertainties in the SCUBA fluxes are 15\%, while the uncertainties in the FTS fluxes are 21\%.
\end{table}

Through calibration with Mars, the observed fluxes of G10.47 and G31.41 agree very well with those recorded by SCUBA; however, G12.21 is much brighter in our observations than the submillimetre continuum measurements indicate, and the values are not within uncertainties. The peak fluxes from the archived SCUBA data for G12.21, which are from two separate observations, agree within uncertainties, suggesting that a variation of the source brightness is unlikely. G12.21 is the faintest source by the SCUBA measurements, and is thus more susceptible to atmospheric subtraction problems than G10.47 and G31.41. Since the largest effect of the variable atmosphere on our observations is to introduce an offset into the overall continuum, and because this offset is cumulative, it is possible that the final flux for this source has been inflated by the bright, variable atmosphere.

\subsection{Molecular line contamination}

The sources observed in this study are hot molecular cores. The moderate temperatures of these objects ($T>100$\,K) cause molecules and elements frozen onto the surface of dust grains during the colder collapse phase to evaporate, releasing these species into the gas phase. The resulting emission lines are often strong in the 350\,GHz spectral window for HMC, and can therefore contaminate measurements of the submillimetre continuum. The sources in this study have been detected in many molecular lines \citep{hatchell98} from multiple species present in the 350\,GHz atmospheric window.

At our low spectral resolution only very bright, relatively broad lines would contribute noticeably to the dust continuum spectrum. Our spectra show little evidence of bright line contamination, as shown in Figure \ref{fig:slopes}, where we have indicated the expected positions of $^{13}$CO (3-2), CS (7-6), CO (3-2), HCN (4-3), and HCO$^+$ (4-3) emission lines. These are the strongest, broadest lines observed in the 350\,GHz window of the hot core G5.89-0.39 \citep{thompson}, which we expect to be similar to the cores studied here. There is some evidence of small peaks in all the spectra, one of which appears to be at the CO(3-2) transition frequency (345.6\ghz); however the data do not allow a conclusive identification. None of the other observed peaks in the spectra are coincident with expected bright line positions. Overall the peaks are small and do not significantly influence the continuum slope, and the contamination of the continuum spectrum by single bright emission lines is therefore not a large problem for this analysis. The concern for this study is then the contamination of the continuum spectrum by a significant number of low-level emission lines that may be present and are not immediately visible, but which may influence the slope of the continuum.

The frequency coverage of molecular line studies for these sources is much less than that covered by SCUBA and the FTS, so no absolute estimates of the level of contamination have been made \citep{hatchell00}. Observing at 350\,GHz with the JCMT, the conversion factor from brightness temperature $T_b$ in Kelvin to flux $S$ in Janskys is approximately $S\,\mbox{(Jy)} \simeq 13 T_b\,\mbox{(K)}$ \citep{hatchell98}. A brightness temperature of 40\,K thus corresponds to $\sim 500$\,Jy.  The line width of strong molecular features, however, is seldom greater than 15\,km\,s$^{-1}$ (at least for the bright core of the line) while the spectral resolution of the FTS for this study is $\sim 1500$\,km\,s$^{-1}$, so that individual lines are diluted by at least a factor of 100. Strong molecular lines, with peak brightness $\sim 40$\,K, will thus add approximately 5\,Jy to the measured flux in a given spectral bin. Therefore, we do not expect contamination of the spectra by individual lines by more than approximately 10-20\%.

The contamination of the submillimetre continuum of hot core sources by molecular lines can range between 10\% and (in extreme cases) 60\% of the total integrated flux \citep{groes94}. \citet{johns03} find that the 350\,GHz continuum emission from their observed protostellar sources, however, is never dominated by molecules other than CO (although line emission from HCN, HNC, CN and CH$_3$OH can contribute a substantial fraction of the contamination for more energetic sources) and line contamination in general is typically less than $\sim10\%$ at 350\,GHz, even for photon-dominated regions. The hot core regions of the sources observed in this study amount to a small fraction of the 350\,GHz beam (1-2\arcsec~in diameter compared with 15\arcsec~FWHM), whereas the more extended emission is from cooler material in which fewer molecules will be excited. Overall, we argue that molecular line contamination of our low resolution 350\,GHz continuum is minimal, and does not significantly affect the analysis. We recognize, however, that it is important to consider the influence of these low-level molecular lines when making any determination of the slope of the dust continuum emission using low spectral resolution submillimetre observations. 

\section{Spectral index results}

\subsection{Calculation of the spectral index}

From Equation \ref{eqn:dust_law}, the dust emission $S(\nu) \propto \nu^\gamma$, where $\gamma=\beta+\alpha$ is the sum of the dust spectral index $\beta$ and the frequency dependence $\alpha$ from the Planck function. For small ranges in frequency $\Delta\,\nu$ around a central frequency $\nu_0$, we can expand this relationship:
\begin{eqnarray} 
S(\nu_0+\Delta\,\nu) &\propto& (\nu_0+\Delta\,\nu)^\gamma \nonumber \\*
&\propto& \nu_0^\gamma \,\biggl(1+\gamma\frac{\Delta\,\nu}{\nu_0}\biggr)
\end{eqnarray}
Thus, for small $\Delta\,\nu\,/\nu_0$, the dust emission is expected to increase linearly with frequency. The FTS passband at 350\,GHz is $\simeq30$\ghz~wide, or $\simeq10$\% of the observing frequency, making the approximation valid. In practice, this window was made slightly smaller due to a decrease in S/N caused by increased levels of contaminating atmospheric flux near the edges of the band, leaving $\Delta\,\nu/\nu\simeq0.08$ on average. A $\chi^2$ minimizing linear fit routine was used to determine the slope of the continuum emission inside the passband. All results quoted here have been determined from the unapodized data to retain the highest spectral resolution. The results varied slightly depending on the fraction of the band used (due to increasing error bars at the band edges when using more of the band, and conversely due to a smaller lever arm when using less of the band); however any variations were within uncertainties. 

Column 2 of Table \ref{tab:betas} shows the best fit continuum emission slope $\gamma$ for each source with uncertainties determined through the $\chi^2$ fitting routine, and Figure \ref{fig:slopes} shows the final spectrum for each source overlaid with the best linear fit. The average reduced $\chi^2$ value for the fits was $\sim0.6$, indicating a good fit to the data within the band. We expect, for a good linear fit, that the reduced $\chi^2$ should be close to unity for a sufficiently large number of data points. The small values of the reduced $\chi^2$ in our fits suggest that the uncertainty per data point has been overestimated by $\sim20$\% for each source. Since the uncertainty in the calculated slope varies as the uncertainty per data point, this implies that the uncertainties in our slopes may be overestimated by $\sim20$\%. 

\begin{table}
\caption{Spectral indices}
\label{tab:betas}
\begin{tabular}{cccc}
\hline
Source & $\gamma_{FTS}$ & $\beta$ & $\gamma_{SCUBA}$ \\
\hline
G10.47 & $3.6\pm1.2$  & $1.6\pm1.2$  & $3.4\pm0.6$  \\
G31.41 & $4.1\pm1.2$  & $2.1\pm1.2$  & $3.3\pm0.6$  \\
G12.21 & $3.3\pm1.1$  & $1.3\pm1.1$  & $4.1\pm0.6$  \\
\hline
\end{tabular}
\end{table}

For emission in the Rayleigh-Jeans limit, $\alpha=2$ and thus the dust spectral index $\beta=\gamma-2$. In order for the R-J approximation to apply at 350\,GHz, $h\nu/kT_d\ll 1$, so $T_d\gg h\nu/k\sim17$\,K. Although HMC are warm, they are surrounded by a shell of cold dust. The observed emission is integrated over the line of sight and the beam, which at 15\arcsec~FWHM is much larger than the hot cores. There is therefore substantial contribution to the observed emission from the colder outer regions along the line of sight. Models of the emission from these sources which take the source temperature and density profiles into account, however, appear to produce spectra which are in the R-J limit at 350\,GHz \citep{hatchell00,hatchell03}. When considering only the peak flux, a greater proportion of the emission from the warm inner regions is contained within the single beam, and the R-J approximation still applies. Column 3 of Table \ref{tab:betas} lists the dust spectral index $\beta$ calculated for each source assuming the 350\,GHz emission is in the R-J limit. We find $\beta$ in the range of $1.3 - 2.1$ with a mean of 1.6.


\subsection{Comparison with SCUBA data}

As discussed in Section 1, the spectral index of dust emission can also be calculated using observations at two widely separated wavelengths. From Equation \ref{eqn:dust_law}, we see that we can solve for $\gamma=\beta+\alpha$ with a ratio of the flux at two different wavelengths:
\begin{equation}
\gamma = \frac{\ln(S_{2}\,/\,S_{1})}{\ln(\nu_2\,/\,\nu_1)} = \beta + \alpha
\label{eqn:scuba_calc}
\end{equation}
For our sources, archived SCUBA data at 450\,$\mu$m (660\,GHz) and 850\,$\mu$m (350\,GHz) are publicly available. In order to properly compare the flux at the two wavelengths, the 8.5\arcsec~resolution 450\,\micron~data were first convolved to the 15\arcsec~resolution of the 850\,\micron~data. The ratio of the peak brightnesses was then used to calculate $\gamma$ from Equation \ref{eqn:scuba_calc}. Column 4 of Table \ref{tab:betas} gives the $\gamma$ value, $\gamma_{SCUBA}$, calculated from the ratio of the 450\,\micron~and 850\,\micron~fluxes for each source. Within uncertainties, $\gamma_{SCUBA}$ is consistent with $\gamma_{FTS}$. The uncertainties in this calculation are dominated by the uncertainty in the source flux at both wavelengths.

If the R-J approximation applies at both frequencies $\nu_1$ and $\nu_2$, then $\alpha=2$ and $\beta=\gamma_{SCUBA}-2$. As shown previously, the emission at 350\,GHz is likely in the R-J limit, but for the R-J approximation to apply at 660\,GHz, $T_d\gg h\nu/k\sim32$\,K.  For the average dust temperatures likely associated with these hot molecular core sources, the 660\,GHz emission cannot be assumed to be in the R-J limit. In this case, less power is contributed to the dust emission from the Planck function, and $\alpha<2$. Using this information, we can use the 350\,GHz (850\,\micron) slope determined using the FTS and $\gamma_{SCUBA}$ to estimate the temperatures of the observed HMC. 

The spectral indices calculated for G10.47 agree well with $\alpha\simeq2$, suggesting that the average dust temperature for this source is high enough ($T_d>60$\,K) for the emission at both wavelengths to be in the R-J limit. G12.21 is also likely warm, while the results for G31.41 indicate a lower average dust temperature. The uncertainties in these temperature estimates are very large due to the large uncertainties in the evaluated slopes; it is clear, however, that with smaller uncertainties in the value of $\beta$ determined from FTS observations, it will be possible to independently determine both $T_d$ and $\beta$ of the dust opacity in star forming regions. 

\subsection{Interpretation}

Our results are consistent with the values of $\beta$ found in embedded \uchII~regions and other HMC. Fitting models of HMC to observational data have found $\beta=1.6-2$ \citep{osorio99,church90}. In a study of 17 \uchII~regions, \citet{hunter98} finds an average spectral index of $\beta=2.0\pm0.25$ through greybody modelling of IRAS, submillimetre and millimetre observations. Observations and modelling of six compact HII regions, one of which was G10.47, by \citet{hoare91} find an average $\beta\simeq1.5$. They note, however, that the frequency dependence is somewhat affected by the assumed density distribution of the model. The dense ridge in Orion has $\beta\simeq1.9$ near embedded infrared sources \citep{gold97}. Overall, the spectral indices of these warmer, likely more evolved star forming regions tend to be higher than those found in cooler, less evolved dense cores, indicating that dust properties have changed through some mechanism which appears to correlate with protostellar age. 

Several dust models predict a dust emissivity frequency dependence similar to that found here, but our observations are not sensitive enough to determine a most likely dust composition. Higher values of $\beta$ have been associated with dust grains covered in thick ice mantles \citep{aanestad,lismenten}, while many dust grain models with combinations of silicate and graphite compositions find a spectral index of $\beta\sim1.5-2$. \citet{ossen94} tabulate the dust opacity properties for nine different dust distributions, including bare grains versus grains with ice mantles (both thick and thin), and various degrees of dust coagulation. The emissivity results presented in this paper are unable to definitively rule out any of the models. The best fit, however, is for bare grains in relatively dense environments, where coagulation is more pronounced (column 2 of Table 1 in \citet{ossen94}).

While we had hoped to determine $\gamma$ to an accuracy of $\pm\,0.1$, the variable atmosphere at 350\,GHz degraded the achieved signal-to-noise ratio by a factor $\ga10$ compared with the instrumental S/N, leading to uncertainties in our calculation of $\beta$ of approximately $\pm1.2$. Even with the atmospheric complications, however, the uncertainties determined here are of the same order as those found when calculating the dust spectral index through multiple wavelength observations of single sources. These results show that with better sky subtraction techniques, it should be possible to determine the spectral index of bright astronomical sources with high accuracy using an FTS. Future instruments with greater sensitivity and precise sky subtraction will enable the measurement of spectral indices of much fainter sources. These results show that careful sky subtraction is \textit{essential} to future dust continuum emission studies with an FTS.

\subsection{An imaging FTS for SCUBA-2}

\begin{table*}
\begin{minipage}{135mm}
\caption{Source brightnesses required to determine $\gamma$ in 12 hours of observing}
\label{tab:ifts}
\begin{tabular}{cccccc}
\hline
Wavelength & Frequency & Resolution & 12 hrs 1$\sigma$ $\Delta$T & Source brightness & Source brightness \\
\micron & \ghz & \ghz & mK & $\gamma \pm 0.1$ (Jy/beam) & $\gamma \pm 0.3$ (Jy/beam) \\
\hline
450 & 660 & 3.0 & 1.44 & 4.7 & 1.6 \\
850 & 350 & 3.0 & 0.2 & 0.40 & 0.14 \\
\hline
\end{tabular}

\medskip
Source brightnesses required to determine the dust spectral index $\gamma$ in 12 hours of observing using an IFTS on SCUBA-2.
\end{minipage}
\end{table*}

SCUBA-2 is the next generation submillimetre instrument currently in development to replace SCUBA in 2006 \citep{holl03}. An imaging FTS (IFTS) is being developed as an ancillary instrument to SCUBA-2 \citep{naylor04}. Based on detector noise calculations alone, the IFTS will be at least 10 times more sensitive per pixel than the single bolometer FTS used in these observations \citep{gom04}. The extremely accurate sky subtraction possible with the IFTS will provide an even more substantial increase in the observational precision. When also considering the variable atmosphere noise, the per-pixel sensitivity of the IFTS increases to at least 100 times greater than in the observations presented in this paper. Using the detector sensitivities in \citet{gom04}, we have calculated that it will be possible to determine in 12 hours of observing with the IFTS the spectral index $\gamma=\beta+\alpha$ of the submillimetre dust emission at 850\,\micron~ to $\pm0.1$ for sources of only 400\,mJy/beam in brightness, and to $\pm0.3$ for sources only $\sim140$\,mJy/beam in brightness. Similar calculations for observations at 450\,\micron~are listed with those at 850\,\micron~in Table \ref{tab:ifts}.

\section{Summary}

We have determined the spectral index of the dust emission of three hot molecular cores solely within the 350\,GHz (850\,\micron) passband using an FTS on the JCMT. We find an average dust spectral index $\beta\simeq1.6$, in agreement with the spectral indices calculated using archived SCUBA data at 450\,\micron~and 850\,\micron. These results for $\beta$ are consistent with measurements of the spectral index in other HMC. The uncertainties in $\beta$ from these observations are dominated by the difficulties in subtracting the very bright and variable submillimetre atmosphere from the data. With better sky subtraction techniques, as will be possible with an imaging FTS on SCUBA-2, these uncertainties will be greatly reduced. The per-pixel sensitivity of the imaging FTS planned for SCUBA-2 will be greater by a factor of $\sim100$ than that of the single pixel detector used in this study, and will allow the determination of the spectral index of dust emission of sources with brightnesses of only a few hundred mJy/JCMT beam. 

\section*{Acknowledgments}
We thank the referee, Jenny Hatchell, for her suggestions which improved this paper. We also thank all TSS involved in FTS runs for their invaluable assistance, as well as G. J. Tompkins, B. Gom and I. Chapman for much FTS-related support. Additional thanks to G. Fuller for providing the source list. RKF also thanks B. Hesman for helpful advice and discussion. This work was funded in part by the University of Victoria (RKF) and by grants from NSERC Canada (DJ, DAN and GRD). The JCMT is operated by the Joint Astronomy Centre in Hilo, Hawaii on behalf of the parent organizations Particle Physics and Astronomy Research Council in the United Kingdom, the National Research Council of Canada and The Netherlands Organization for Scientific Research. SCUBA-2 is a jointly funded project through the JCMT Development Fund with substantial additional contributions from the UK Office of Science and Technology and the Canada Foundation for Innovation.

\label{lastpage}


\begin{thebibliography}{99}
\bibitem[\protect\citeauthoryear{Aanestad}{1975}]{aanestad} Aanestad, P.A. 1975, ApJ, 200, 30
\bibitem[\protect\citeauthoryear{Archibald et al.}{2002}]{arch02} Archibald, E.N., Jenness, T., Holland, W.S. et al. 2002, MNRAS, 336, 1
\bibitem[\protect\citeauthoryear{Beuther, Schilke \& Wyrowski}{2004}] {beuther04} Beuther, H., Schilke, P. \& Wyrowski, F. 2004, ApJ, in press
\bibitem[\protect\citeauthoryear{Chapman}{2001}]{chapman} Chapman, I. MSc thesis, Lethbridge, 2001
\bibitem[\protect\citeauthoryear{Chapman, Naylor \& Phillips}{2004}]{chapman04} Chapman, I.M., Naylor, D.A. \& Phillips, R.R. 2004, MNRAS, 354, 621
\bibitem[\protect\citeauthoryear{Churchwell, Wolfire \& Wood}{1990}]{church90} Churchwell, E., Wolfire, M.G. \& Wood, D.O.S. 1990, ApJ, 354, 247
\bibitem[\protect\citeauthoryear{Davis et al.}{1997}]{davis97} Davis, G.R., Naylor, D.A., Griffin, M.J., Clark, T.A. \& Holland, W.S. 1997, {\it Icarus}, 130, 387
\bibitem[\protect\citeauthoryear{Draine \& Lee}{1984}]{draine84} Draine, B.T. \& Lee, H.M. 1984, ApJ, 285, 89
\bibitem[\protect\citeauthoryear{Goldsmith, Bergin \& Lis}{1997}]{gold97} Goldsmith, P.F., Bergin, E.A., \& Lis, D.C. 1997, ApJ, 491, 615
\bibitem[\protect\citeauthoryear{Gom \& Naylor}{2004}]{gom04} Gom, B.G. \& Naylor, D.A. 2004. in Proc. SPIE, Vol. 5498, Millimeter and Submillimeter Detectors for Astronomy IIA, ed. Zmuidzinas, J., Holland, W.S., Withington, S. (SPIE) 695
\bibitem[\protect\citeauthoryear{Groesbeck, Phillips \& Blake}{1994}]{groes94} Groesbeck, T.D., Phillips, T.G. \& Blake, G.A. 1994, ApJs , 94, 147
\bibitem[\protect\citeauthoryear{Hatchell et al.}{1998}]{hatchell98} Hatchell, J., Thompson, M.A., Millar, T.J. \& Macdonald, G.H. 1998, A\&AS, 133, 29
\bibitem[\protect\citeauthoryear{Hatchell et al.}{2000}]{hatchell00} Hatchell, J., Fuller, G.A., Millar, T.J., Thompson, M.A., Macdonald, G.H. 2000, A\&A. 357, 637
\bibitem[\protect\citeauthoryear{Hatchell \& van der Tak}{2003}]{hatchell03} Hatchell, J. \& van der Tak, F.F.S. 2003, A\&A, 409, 589
\bibitem[\protect\citeauthoryear{Hildebrand}{1983}]{hild83} Hildebrand, R.H. 1983, QJRAS, 24, 267
\bibitem[\protect\citeauthoryear{Hoare, Roche \& Glencross}{1991}]{hoare91} Hoare, M.G., Roche, P.F. \& Glencross, W.M. 1991, MNRAS, 251, 584
\bibitem[\protect\citeauthoryear{Hogerheijde \& Sandell}{2000}]{hoger00} Hogerheijde, M.R. \& Sandell, G. 2000, ApJ, 534, 880
\bibitem[\protect\citeauthoryear{Holland et al.}{2003}]{holl03} Holland, W.S., Duncan, W.D., Kelly, B.D., Irwin, K.D., Walton, A.J., Ade, P.A.R., Robson, E.I. 2003, in Proc. SPIE, Vol. 4855, Millimeter and Submillimeter Detectors for Astronomy, ed. Phillips, T.G., Zmuidzina, J. (SPIE) 1
\bibitem[\protect\citeauthoryear{Hunter}{1998}]{hunter98} Hunter, T.R. 1998, PASP, 110, 634
\bibitem[\protect\citeauthoryear{Johnstone, Boonman \& van Dishoeck}{2003}]{johns03} Johnstone, D., Boonman, A.M.S. \& van Dishoeck, E.F. 2003, A\&A, 412, 157
\bibitem[\protect\citeauthoryear{Kurtz et al.}{2000}]{kurtz00} Kurtz, S., Cesaroni, R., Churchwell, E., Hofner, P., \& Walmsley, C.M. 2000, in Protostars \& Planets IV, ed. V. Mannings, A.P. Boss, \& S.S. Russell (Tucson: Univ. Arizona Press)
\bibitem[\protect\citeauthoryear{Lis \& Menten}{1998}]{lismenten} Lis, D.C. \& Menten, K.M. 1998, ApJ, 507, 794
\bibitem[\protect\citeauthoryear{Naylor et al.}{1994}]{naylor94} Naylor, D.A., Davis, G.R., Griffin, M.J., Clark, T.A., Gautier, D. \& Marten, A. 1994, A\&A, 291, L51
\bibitem[\protect\citeauthoryear{Naylor et al.}{2000}]{naylor00} Naylor, D.A., Davis, G.R., Gom, B.G., Clark, T.A., Griffin, M.J. 2000, MNRAS, 315, 622
\bibitem[\protect\citeauthoryear{Naylor et al.}{2003}]{naylor03} Naylor, D.A., Gom, B.G., Schofield, I.S., Tompkins, G.J., Davis, G.R. 2003, Proc. SPIE, in Proc. SPIE, Vol. 4855, Millimeter and Submillimeter Detectors for Astronomy, ed. Phillips, T.G., Zmuidzina, J. (SPIE) 540
\bibitem[\protect\citeauthoryear{Naylor \& Gom}{2004}]{naylor04} Naylor, D.A. \& Gom, B.G. 2004, in Proc. SPIE, Vol. 5159, Imaging Spectrometry IX, ed. Shen, S.S., Lewis, P.E. (SPIE) 91
\bibitem[\protect\citeauthoryear{Naylor et al.}{1999}]{naylor99} Naylor, D.A., Gom, B. G., Ade, P. A. R., Davis, J. E. 1999, Rev. Sci. Instruments, 70, 4097
\bibitem[\protect\citeauthoryear{Ossenkopf \& Henning}{1994}] {ossen94} Ossenkopf, V. \& Henning, Th. 1994, A\&A, 291, 943
\bibitem[\protect\citeauthoryear{Osorio, Lizano \& D'Alessio}{1999}]{osorio99} Osorio, M., Lizano, S. \& D'Alessio, P. 1999, ApJ, 525, 808
\bibitem[\protect\citeauthoryear{Pollack et al.}{1994}]{poll94} Pollack, J.B., Hollenbach, D., Beckwith, S., Simonelli, D.P., Roush, T. \& Fong, W. 1994, ApJ, 421, 615
\bibitem[\protect\citeauthoryear{Privett, Jenness \& Matthews}{1998}]{starlink} Privett G., Jenness T., Matthews H., 1998. Starlink User Note 213.2
\bibitem[\protect\citeauthoryear{Serabyn \& Weisstein}{1995}]{serabyn} Serabyn, E. \& Weisstein, E.W. 1995, ApJ, 451, 238
\bibitem[\protect\citeauthoryear{Thompson \& Macdonald}{1999}]{thompson} Thompson, M.A. \& Macdonald, G.H. 1999, A\&AS, 135, 531
\bibitem[\protect\citeauthoryear{Visser et al.}{1998}]{visser98} Visser, A.E., Richer, J.S., Chandler, C.J. \& Padman, R. 1998, MNRAS, 301, 585
\bibitem[\protect\citeauthoryear{Walsh et al.}{2003}]{walsh03} Walsh, A.J., Macdonald, G.H., Alvey, N.D.S., Burton, M.G., Lee, J.K. 2003, A\&A, 410, 597
\bibitem[\protect\citeauthoryear{Williams, Fuller \& Sridharan}{2004}] {will04} Williams, S.J., Fuller, G.A. \& Sridharan, T.K. 2004, A\&A, 417, 115
\bibitem[\protect\citeauthoryear{Wood \& Churchwell}{1989}]{wc89} Wood, D.O.S. \& Churchwell, E. 1989, ApJS, 69, 831
\end{thebibliography}
\end{document}